\documentclass[11pt]{article}
\usepackage{amssymb,latexsym}
\def\lb{[\![}
\def\rb{]\!]}
\def\al{\alpha}

\def\de{\delta}
\def\ep{\epsilon}

\def\la{\lambda}

\def\si{\sigma}

\def\La{\Lambda}

\def\Z{\mathbb{Z}}
\def\beq{\begin{equation}}
\def\eeq{\end{equation}}
\def\bea{\begin{eqnarray}}
\def\eea{\end{eqnarray}}
\def\beas{\begin{eqnarray*}}
\def\eeas{\end{eqnarray*}}
\def\nn{\nonumber}

\begin{document}

\begin{center}
{\bf \large Realizations of the Lie superalgebra $q(2)$ and 
applications}\\[5mm]
{\bf N.\ Debergh}\footnote{Email: Nathalie.Debergh@ulg.ac.be, Chercheur IISN} \\
Physique Th\'eorique Fondamentale,\\
Institut de Physique, B5,\\
Universit\'e de Li\`ege, Sart Tilman, B-4000 Li\`ege, Belgium,\\[2mm]
{\bf and} \\ {\bf J.\ Van der Jeugt}\footnote{Email: Joris.VanderJeugt@rug.ac.be}\\
Department of Applied Mathematics and Computer Science,\\
University of Ghent, Krijgslaan 281-S9, B-9000 Gent, Belgium.
\end{center}

\vskip 20mm

\begin{abstract}
The Lie superalgebra $q(2)$ and its class of irreducible representations
$V_p$ of dimension~$2p$ ($p$ being a positive integer) are considered.
The action of the $q(2)$ generators on a basis of $V_p$ is given
explicitly, and from here two realizations of $q(2)$ are determined.
The $q(2)$ generators are realized as differential operators
in one variable $x$, and the basis vectors of $V_p$ as 2-arrays
of polynomials in $x$. Following such realizations, it is observed
that the Hamiltonian of certain physical models can be written
in terms of the $q(2)$ generators. In particular, the models given
here as an example are the sphaleron model, the Moszkowski model
and the Jaynes-Cummings model. For each of these, it is shown
how the $q(2)$ realization of the Hamiltonian is helpful in
determining the spectrum.
\end{abstract}

\section{Introduction}

Since their introduction in supersymmetry~\cite{Golfand,Wess,Salam},
Lie superalgebras and their irreducible representations 
(simple modules) have been the subject of much attention in both 
the mathematical~\cite{Kac77, Kac78, Scheunert} and the physics
literature, where both finite dimensional~\cite{Corwin, Balantekin, Hurni}
and infinite dimensional 
representations~\cite{Heidenreich,Freedman,Flato,Dobrev,VanderJeugt}
have been studied.
When Kac obtained his classification~\cite{Kac77}
of simple Lie superalgebras, he subdivided them into 
the classical Lie superalgebras and the Lie superalgebras
of Cartan type.
The classical Lie superalgebras consist of the basic
Lie superalgebras -- $A(m,n)$, $B(m,n)$, $C(n)$, $D(m,n)$ and the 
exceptional ones $D(2,1;\al)$, $G(3)$ and $F(4)$ --
and the strange series $P(n)$ and $Q(n)$. 
The basic Lie superalgebras have made their appearance 
in various physical models. As far as we know, the
strange Lie superalgebras have not been used in relation to
any physical model or example. 
In this paper, we shall discuss the strange Lie superalgebra
$Q(1)$ of rank~1; more precisely we shall be dealing with its
central extension which is usually denoted by $q(2)$~\cite{Penkov}.
It will be shown that $q(2)$ has a class of interesting
representations $V_p$ labelled by a positive integer $p$.
These representations allow for certain realizations of $q(2)$,
and it will be shown that these realizations in turn are
appropriate for the study of certain physical models~:
the so-called sphaleron model, the Moszkowski model,
and the Jaynes-Cummings model.

The strange Lie superalgebras $q(n)$ can be considered as a
super-analogue of $gl(n)$. Representations of $q(n)$ have
been studied from the mathematical point of view.
In~\cite{Penkov, PS1, PS2}, the finite dimensional irreducible
{\em graded} representations of $q(n)$ have been determined
together with their characters, both in the so-called typical
and atypical case. These representations possess the strange 
property that the multiplicity of the highest weight
is in general greater than~1~\cite{PS1}. 
More recently, a new class of finite dimensional
irreducible representations of $q(n)$ was
determined~\cite{PV}. These representations are {\em not graded}
and thus they are not among the ones classified by
Penkov and Serganova~\cite{PS1}. However, they possess many
other interesting properties~: the highest weight has 
multiplicity~1, they can be equipped with an inner product,
and in an apropriate context they can be considered as 
Fock spaces.

In the present paper we shall concentrate on these representations
for the Lie superalgebra $q(2)$. 
The representations $V_p$ are of dimension $2p$ ($p$ is
a positive integer). When decomposed to the even
subalgebra $gl(2)$ of $q(2)$, $V_p$ consists of the
direct sum of two $gl(2)$ irreps~: one of dimension
$p+1$ and one of dimension $p-1$. Having two $gl(2)$
irreps of such dimension as part of an irreducible
representation of another algebra (namely $q(2)$),
will help in determining physical applications for
the representations $V_p$.

The structure of the paper is as follows.
In section~2 the algebra $q(2)$ and its class of
representations $V_p$ are defined. 
In section~3 we shall discuss a relation between
these representations and certain representations
of $so(4)$. Two realizations of $q(2)$ and of the
corresponding representations $V_p$ will be given
in section~4. 
The appearance and usefullness of these realizations
in physical models will then be illustrated in the
following sections~: the sphaleron model in section~5,
the Moszkowski model in section~6  
and the Jaynes-Cummings model in section~7.

\section{The Lie superalgebra $q(2)$ and the representations $V_p$}

For the definition of $q(n)$ and a corresponding class
of representations, we refer to~\cite{PV}. Here we
shall deal only with the case $n=2$.
The Lie superalgebra $q(2)$ has a basis consisting of 4
even elements $e_{ij}^{\bar 0}$ ($i,j=0,1$) and 4
odd elements $e_{ij}^{\bar 1}$ ($i,j=0,1$), 
satisfying the bracket relation
\beq
\lb e_{ij}^\si, e_{kl}^\theta \rb = \de_{jk} e_{il}^{\si+\theta} -
(-1)^{\si\theta} \de_{il} e_{kj}^{\si+\theta},
\eeq
where $\si,\theta\in\Z_2=\{\bar 0, \bar 1\}$, and $i,j,k,l\in\{0,1\}$. 
Here, $\lb\,,\,\rb$ stands for the
Lie superalgebra bracket, which could be a commutator or
an anti-commutator, depending on the grading of the 
elements considered. We write explicitly $[\,,\,]$ (
resp.\ $\{\,,\,\}$) if this
stands for a commutator (resp.\ anti-commutator).

It is clear that the even part of $q(2)$ (i.e.\ the 4 elements
with upper index equal to $\bar 0$) is the Lie algebra $gl(2)$.
For convenience, a different notation will be introduced
for the root vectors, i.e.\ the elements $e_{ij}^\si$
with $i\ne j$, since these elements can be interpreted
as ``creation and annihilation operators'' for $q(2)$~\cite{PV}.
So we put~:
\bea
&& b^+=e_{10}^{\bar 0},\qquad b^-=e_{01}^{\bar 0}, \\
&& f^+=e_{10}^{\bar 1},\qquad f^-=e_{01}^{\bar 1}.
\eea
These operators satisfy certain triple relations 
(see~\cite[(8)--(11)]{PV}), and together with their 
supercommutators they form a basis of $q(2)$.

The algebra $q(2)$ has finite dimensional representations labelled by
a positive integer $p$. 
The representation space $V_p$ arises as a quotient
module $V_p=\bar V_p / M_p$ of an infinite dimensional $q(2)$
module $\bar V_p$ by its maximal submodule $M_p$~\cite{PV}. 
The space $\bar V_p$ is spanned by the vectors
\bea
&&v_k= (b^+)^k v_0,\; k=0,1,\ldots;\nn\\ 
&&w_k= (b^+)^{k-1} f^+ v_0,\; k=1,2,\ldots,
\eea
where $v_0$ is a vacuum (or highest weight vector) satisfying~:
\bea
&& e_{00}^{\bar 0} v_0 = p v_0, \qquad
e_{00}^{\bar 1} v_0 = \sqrt{p} v_0,\nn \\ 
&& e_{11}^{\bar 0} v_0 = 0, \qquad
e_{11}^{\bar 1} v_0 = 0,\\
&& b^- v_0 = f^- v_0 =0. \nn
\eea
The following actions in $\bar V_p$ of the creation and annihilation 
operators on $v_k$ and $w_k$ can be computed~:
\bea
&& b^+ v_k = v_{k+1},\qquad b^+ w_k= w_{k+1}, \nn\\
&& f^+ v_k = w_{k+1},\qquad f^+ w_k=0, \nn\\
&& b^- v_k = k(p-k+1) v_{k-1}, \nn\\
&& f^- v_k = k \sqrt{p}\; v_{k-1} -k(k-1) w_{k-1}, \nn\\
&& b^- w_k = \sqrt{p}\; v_{k-1} + (k-1)(p-k) w_{k-1}, \nn\\
&& f^- w_k = p v_{k-1} - (k-1)\sqrt{p}\; w_{k-1}. \label{q24}
\eea
In $\bar V_p$, $v_p-\sqrt{p}\; w_p$ is a primitive vector (the
action of $b^-$ and $f^-$ on it are zero) generating
the submodule $M_p$. The quotient module $V_p=\bar V_p/M_p$ is therefore
a finite dimensional module. A set of basis vectors of $V_p$,
together with the corresponding weight
in the natural basis $(\ep_0, \ep_1)$ of the
$gl(2)$ weight space, is given by
\beq
\begin{array}{lll}
v_0 && p\ep_0\\
v_1, w_1 && (p-1)\ep_0+\ep_1\\
v_2, w_2 && (p-2)\ep_0+2\ep_1\\
\vdots && \vdots \\
v_{p-1}, w_{p-1} && \ep_0+(p-1)\ep_1\\
v_p+\sqrt{p}\; w_p && p\ep_1.
\end{array}
\label{vwbasis}
\eeq
The top and bottom weight appear with multiplicity~1, 
the other weights have multiplicity~2. 
Observe that we use the same notation for vectors in
$V_p$ and in $\bar V_p$.

{}From the above weight structure one can determine the
decomposition of this finite dimensional $q(2)$ module with respect
to the even subalgebra $gl(2)\subset q(2)$~:
\beq
V_p \rightarrow (p,0) \oplus (p-1,1),\qquad\qquad (p>1).
\eeq
So $V_p$ splits into two irreducible $gl(2)$ modules, 
both of which have been labelled by their highest weight 
(in the $(\ep_0,\ep_1)$-basis). 
In other words, the two components of the $gl(2)$
representations have dimension $p+1$ and $p-1$;
often this $gl(2)$ representation would be denoted
by ${\cal D}^{({p\over 2})} \oplus {\cal D}^{({p\over 2}-1)}$.

The actions of the remaining $q(2)$ basis elements on the
representation space $V_p$ can easily be determined~:
\beq
\begin{array}{ll}
e^{\bar 0}_{00} v_k=(p-k) v_k, & e^{\bar 0}_{00} w_k=(p-k) w_k,\\
e^{\bar 0}_{11} v_k=k\; v_k, & e^{\bar 0}_{11} w_k=k\; w_k,\\
e^{\bar 1}_{00} v_k=\sqrt{p} v_k-k\; w_k, &
 e^{\bar 1}_{00} w_k=v_k-\sqrt{p}w_k,\\
e^{\bar 1}_{11} v_k=k\; w_k, &
e^{\bar 1}_{11} w_k=v_k.
\end{array}
\eeq

On the representation space $V_p$, a positive-definite
metric can be introduced by requiring
\beq
\langle v_0| v_0 \rangle =1, \qquad
\langle b^+ v| v' \rangle = \langle v| b^- v' \rangle, \qquad
\langle f^+ v| v' \rangle = \langle v| f^- v' \rangle, \qquad
\forall v,v'\in V_p.
\eeq
Then
\beq
\langle v_k| v_l \rangle =\de_{kl}{k!p!\over (p-k)!}, \qquad
\langle w_k| w_l \rangle =\de_{kl}{(k-1)!p!\over (p-k)!}, \qquad
\langle v_k| w_l \rangle =\de_{kl}{k!p!\over (p-k)!\sqrt{p}}.
\eeq
Because of the last relation, the basis~(\ref{vwbasis}) 
is not orthogonal with respect to 
this metric, so it will be convenient to introduce
another (and more convenient) orthogonal basis of $V_p$ 
as follows~:
\bea
\La_k &=& {(p-k)!\over p!}\; v_k,\qquad (k=0,1,\ldots,p-1),
\label{La_k}\\
\La_p &=& {1\over 2 p!} (v_p+\sqrt{p}\, w_p), \label{La_p}\\
\chi_l &=& {(p-l-1)!\over p!} (v_l-\sqrt{p}\, w_l),
\qquad (l=1,2,\ldots,p-1). \label{chi_l}
\eea
The action of the creation and annihilation operators on this basis
reads (in the following equations, $k=0,1,\ldots,p$ and
$l=1,2,\ldots,p-1$)~:
\bea
b^- \La_k &=& k \La_{k-1}, \nn\\
b^- \chi_l &=& (l-1) \chi_{l-1}, \nn\\
b^+ \La_k &=& (p-k) \La_{k+1}, \nn\\
b^+ \chi_l &=& (p-l-1) \chi_{l+1}, \nn\\
f^- \La_k &=& (k \La_{k-1} + k(k-1) \chi_{k-1})/\sqrt{p}, \nn\\
f^- \chi_l &=& -(\La_{l-1} + (l-1)\chi_{l-1})/\sqrt{p}, \nn\\
f^+ \La_k &=& ((p-k)\La_{k+1}-(p-k)(p-k-1)\chi_{k+1})/\sqrt{p}, \nn\\
f^+ \chi_l &=& (\La_{l+1} -(p-l-1)\chi_{l+1})/\sqrt{p}. 
\label{bLachi}
\eea
Note that in all computations, one has to remember to work in the
quotient module $V_p=\bar V_p/M_p$, where $M_p$ is generated by the
primitive vector $v_p-\sqrt{p} w_p$ of $\bar V_p$. This often
requires a separate calculation for the cases $k=p$ or $k=p-1$.
For example,
\[
b^+ \La_{p-1} = {1\over p!} v_p =
{1\over p!} \left( v_p -{1\over 2}(v_p-\sqrt{p} w_p)\right) =
{1\over 2 p!} (v_p+\sqrt{p} w_p)= \La_p.
\]
The actions of the remaining $q(2)$ elements
in this basis are given by
\bea
e^{\bar 0}_{00} \La_k &=& (p-k)\La_k, \nn\\
e^{\bar 0}_{00} \chi_l &=& (p-l)\chi_l, \nn\\
e^{\bar 0}_{11} \La_k &=& k\,\La_k, \nn\\
e^{\bar 0}_{11} \chi_l &=& l\,\chi_l, \nn\\
e^{\bar 1}_{00} \La_k &=& ((p-k)\La_k + k(p-k)\chi_k)/\sqrt{p}, \nn\\
e^{\bar 1}_{00} \chi_l &=& (\La_l -(p-l) \chi_l)/\sqrt{p}, \nn\\
e^{\bar 1}_{11} \La_k &=& (k\La_k - k(p-k)\chi_k)/\sqrt{p}, \nn\\
e^{\bar 1}_{11} \chi_l &=& -(\La_l + l\chi_l)/\sqrt{p},
\label{eLachi}
\eea
where again $k=0,1,\ldots,p$ and
$l=1,2,\ldots,p-1$.
Observe that the subalgebra $gl(2)$ with basis $\{ b^+, b^-, 
e^{\bar 0}_{00}, e^{\bar 0}_{11} \}$ acts irreducibly on the 
vectors $\La_k$ ($k=0,1,\ldots,p$) and $\chi_l$
($l=1,2,\ldots,p-1$); so from here the decomposition of
$V_p$ into two irreducible $gl(2)$ irreps is obvious.

\section{A relation with $so(4)$ representations}

Consider the Lie algebra $so(4) \equiv sl(2)\oplus sl(2)$
with generators $J_i$ and $K_i$ ($i=0,\pm$) and 
commutation relations~:
\bea
&& [J_0,J_\pm] = \pm J_\pm, \qquad [J_+,J_-]=2J_0, \nn\\
&& [K_0,K_\pm] = \pm K_\pm, \qquad [K_+,K_-]=K_0, \label{so4}\\
&& [J_i,K_j]=0. \nn
\eea
Rather than dealing with the abstract generators of $so(4)$,
we shall consider these generators in a particular representation.
The operators $J_i$ ($i=0,\pm$) are realized in the
representation ${\cal D}^{({p-1\over 2})}$ of $sl(2)$ (with $p$
a positive integer), and the operators $K_i$ ($i=0,\pm$) 
are realized in the representation ${\cal D}^{({1\over 2})}$ 
of $sl(2)$. We shall continue to denote the representatives of
the abstract operators~(\ref{so4}) by the same names, 
$J_i$ and $K_i$. Thus the $K_i$ satisfy
\beq
(K_\pm)^2=0,\qquad K_0^2={1\over 4}I,\qquad
\{K_+,K_-\}=I, \qquad \{K_0,K_\pm\} = 0,
\eeq
where $I$ is the identity operator.

The Lie algebra $so(4)=sl(2)\oplus sl(2)$ has the subalgebra $sl(2)$
with generators $J_i +K_i$ ($i=0,\pm$).
Since in the present realization the tensor product 
${\cal D}^{({p-1\over 2})} \otimes {\cal D}^{({1\over 2})}$ 
decomposes as ${\cal D}^{({p\over 2})} \oplus
{\cal D}^{({p\over 2}-1)}$, the representation of $so(4)$
considered here decomposes as ${\cal D}^{({p\over 2})} \oplus
{\cal D}^{({p\over 2}-1)}$ with respect to this $sl(2)$
subalgebra. This implies that the $so(4)$ representation
space is isomorphic to the space $V_p$, with the same 
$sl(2)$ action. Denoting the representatives of $q(2)$
in $V_p$ again by $b^\pm$, $f^\pm$, $e^\si_{ii}$ 
($\si=\bar 0, \bar 1$, $i=0,1$), the following identification
holds~:
\bea
&& b^-=J_+ + K_+,\qquad b^+=J_- + K_-,\qquad 
e_{00}^{\bar 0} - e_{11}^{\bar 0} = 2J_0+2K_0, \nn\\
&& f^-=\sqrt{p} K_+, \qquad f^+ = \sqrt{p} K_-, \qquad
e_{00}^{\bar 1} - e_{11}^{\bar 1} = 2\sqrt{p}K_0, 
\label{bJK}\\
&& e_{00}^{\bar 0} + e_{11}^{\bar 0} =pI,\qquad
e_{00}^{\bar 1} + e_{11}^{\bar 1} = {2\over\sqrt{p}}
(2J_0K_0 + J_+K_- +J_-K_+ +{1\over 2}). \nn
\eea
These relations can be verified by considering the 
representations of the $so(4)$ generators in a standard
basis of ${\cal D}({p-1\over 2},{1\over 2}) =
{\cal D}^{({p-1\over 2})} \otimes {\cal D}^{({1\over 2})}$,
and comparing with~(\ref{bLachi})-(\ref{eLachi}).
Indeed, let the standard basis of ${\cal D}({p-1\over 2},{1\over 2})$
be given by
\[
|{p-1\over 2},m\rangle \otimes |{1\over 2},\mu\rangle,
\]
where $m=-{p-1\over 2},-{p-1\over 2}+1,\ldots,{p-1\over 2}$
and $\mu=\pm{1\over 2}$, then the standard action of the
$so(4)$ basis elements reads
\bea
J_0 \;|{p-1\over 2},m\rangle \otimes |{1\over 2},\mu\rangle &=&
m \;|{p-1\over 2},m\rangle \otimes |{1\over 2},\mu\rangle, \nn\\
J_{\pm}\; |{p-1\over 2},m\rangle \otimes |{1\over 2},\mu\rangle &=&
\left( ({p-1\over 2}\mp m)({p-1\over 2}\pm m+1) \right)^{1/2}
|{p-1\over 2},m\pm 1\rangle \otimes |{1\over 2},\mu\rangle, \nn\\
K_0\; |{p-1\over 2},m\rangle \otimes |{1\over 2},\mu\rangle &=&
\mu\; |{p-1\over 2},m\rangle \otimes |{1\over 2},\mu\rangle, \nn\\
K_{\pm}\; |{p-1\over 2},m\rangle \otimes |{1\over 2},\mu\rangle &=&
\left( ({1\over 2}\mp \mu)({1\over 2}\pm \mu+1) \right)^{1/2}
|{p-1\over 2},m\rangle \otimes |{1\over 2},\mu\pm 1\rangle .
\label{JKaction}
\eea
Using the following relation between the $(\La_k, \chi_l)$-basis
and the present one,
\bea
\La_k&=& \sqrt{(p-k)!k!\over p!}\left( \sqrt{p-k\over p}\ 
|{p-1\over 2},{p-1\over 2}-k\rangle \otimes |{1\over 2},{1\over 2}\rangle
\right. \\
&& \qquad \left. +\sqrt{k\over p}\ 
|{p-1\over 2},{p+1\over 2}-k\rangle \otimes 
|{1\over 2},-{1\over 2}\rangle \right),\\
\chi_l&=& \sqrt{(p-l-1)!(l-1)!\over p!}\left( \sqrt{l\over p}\ 
|{p-1\over 2},{p-1\over 2}-l\rangle \otimes |{1\over 2},{1\over 2}\rangle
\right. \\
&& \qquad\left. -\sqrt{p-l\over p}\ 
|{p-1\over 2},{p+1\over 2}-l\rangle \otimes 
|{1\over 2},-{1\over 2}\rangle \right),
\eea
it is straightforward to verify that~(\ref{bJK}) holds, using the
actions~(\ref{bLachi})-(\ref{eLachi}) and~(\ref{JKaction}).

Observe that $so(4)$ has 2 Casimir operators $C_1$ and $C_2$,
which are independent in general~: 
\bea
C_1&=& J_0^2+ K_0^2 +{1\over 2} \{J_+,J_-\} +{1\over 2} \{K_+,K_-\}\\
C_2&=& J_0^2- K_0^2 +{1\over 2} \{J_+,J_-\} -{1\over 2} \{K_+,K_-\}.
\eea
In the present representation, however, these operators are not
independent. They can be rewritten in terms of the $q(2)$
operators, in which case $C_1$ and $C_2$ coincide apart from
a multiple of the operator $e_{00}^{\bar 0} + e_{11}^{\bar 0}$
(with eigenvalue $p$ in the representation). The Casimirs $C_1$
and $C_2$ have the value $2p^2-1$ and $2p^2-4$ respectively.

\section{Two realizations of $q(2)$ and its representation $V_p$}

In order to find applications of the algebra $q(2)$ and its
representations $V_p$, it will be useful to construct certain
differential realizations of $q(2)$. Here we shall give two different
differential realizations. The main difference comes from the 
distinction between the spaces of polynomials that the $q(2)$
elements act upon. 

A simple realization of $q(2)$ is found by realizing the
basis elements $\La_k$, $\chi_l$ as follows~:
\beq
\La_k=\left(\begin{array}{c} x^k \\ 0 \end{array}\right),
k=0,1,\ldots,p, \qquad
\chi_l=\left(\begin{array}{c} 0 \\ x^{l-1} \end{array}\right),
l=1,2,\ldots,p-1. \label{Lachi-x}
\eeq
Thus the basis elements are $(2\times 1)$-arrays of polynomials
in a variable $x$. The representation space can then be
identified with
\beq
\left(\begin{array}{c} {\cal P}(p) \\ {\cal P}(p-2) \end{array}\right),
\label{Pp}
\eeq
where ${\cal P}(m)$ stands for the space of polynomials in $x$
of degree at most $m$, thus ${\cal P}(m)$ has a basis
$\{ 1,x,\ldots, x^m\}$. The Lie superalgebra $q(2)$ will 
have a realization preserving the space~(\ref{Pp}).

With this realization of the basis vectors $\La_k$ and $\chi_l$,
a differential realization for $q(2)$ is easily derived 
from~(\ref{bLachi})-(\ref{eLachi}). There comes~:
\bea
&&b^-=\frac{d}{dx}, \qquad \qquad
b^+=-x^2\frac{d}{dx}+(p-1)x+x\si_3, \nn\\
&&e^{\bar 0}_{00}-e^{\bar 0}_{11}=-2x\frac{d}{dx}+p-1+\si_3, \qquad
\qquad e^{\bar 0}_{00}+e^{\bar 0}_{11}=p, \nn\\
&&f^-=\frac{1}{\sqrt{p}}(\frac{d}{dx}\si_3 - \si_+ + 
\frac{d^2}{dx^2}\si_-), \nn\\
&&f^+=\frac{1}{\sqrt{p}}(-x^2\frac{d}{dx}+(p-1)x)\si_3 + 
\frac{1}{\sqrt{p}}x + \frac{1}{\sqrt{p}}x^2\si_+ \nn\\
&&\qquad - \frac{1}{\sqrt{p}}(x^2\frac{d^2}{dx^2}+
2(1-p)x\frac{d}{dx}+p(p-1))\si_-, \nn\\
&&e^{\bar 1}_{00}-e^{\bar 1}_{11}=\frac{1}{\sqrt{p}}
(-2x\frac{d}{dx}+p-1)\si_3 + \frac{1}{\sqrt{p}} + 
\frac{2}{\sqrt{p}}x\si_+ 
 + \frac{2}{\sqrt{p}}(-x\frac{d^2}{dx^2}+
(p-1)\frac{d}{dx})\si_- ,\nn\\
&&e^{\bar 1}_{00}+ e^{\bar 1}_{11} = \sqrt{p} \si_3.
\label{real1}
\eea
Herein, $\si_\pm$ and $\si_3$ are the common notations
for the Pauli matrices. We shall refer to~(\ref{real1}) as
the first differential realization of $q(2)$.

A second useful realization of $q(2)$ will be found by 
considering a different basis for $V_p$.
Let, for $k=0,1,\ldots, p-1$,
\bea
\mu_k &=& \La_{p-k} -k \chi_{p-k}, \label{mu1}\\
\mu_{p+k} &=& \La_{p-k-1} +(p-k-1)\chi_{p-k-1}. \label{mu2}
\eea
Then the action of the $q(2)$ operators on this new basis
reads~:
\bea
&& b^+ \mu_k = k \mu_{k-1}, \qquad
b^+ \mu_{p+k} = \mu_k+ k\mu_{p+k-1}, \nn\\
&& f^+ \mu_k=0, \qquad f^+\mu_{p+k}= \sqrt{p} \mu_k, \nn\\
&& b^-\mu_k = (p-k-1)\mu_{k+1} + \mu_{p+k}, \qquad
b^-\mu_{p+k} = (p-k-1)\mu_{p+k+1}, \nn\\
&& f^- \mu_k = \sqrt{p} \mu_{p+k}, \qquad
f^- \mu_{p+k} =0, \nn\\
&& (e_{00}^{\bar 0} + e_{11}^{\bar 0})\mu_k= p\mu_k, \qquad
(e_{00}^{\bar 0} + e_{11}^{\bar 0})\mu_{p+k}= p\mu_{p+k}, \nn\\
&& (e_{00}^{\bar 0} - e_{11}^{\bar 0})\mu_k= (2k-p)\mu_k, \qquad
(e_{00}^{\bar 0} - e_{11}^{\bar 0})\mu_{p+k}= (2k+2-p)\mu_{p+k}, \nn\\
&& (e_{00}^{\bar 1} + e_{11}^{\bar 1})\mu_k= {1\over\sqrt{p}}
(p-2k)\mu_k+ {1\over\sqrt{p}}(2k)\mu_{p+k-1}, \nn\\
&& (e_{00}^{\bar 1} + e_{11}^{\bar 1})\mu_{p+k}= {1\over\sqrt{p}}
(2k+2-p)\mu_{p+k}+ {2\over\sqrt{p}}(p-k-1)\mu_{k+1}, \nn\\
&& (e_{00}^{\bar 1} - e_{11}^{\bar 1})\mu_k= -\sqrt{p}\mu_k, \qquad 
(e_{00}^{\bar 1} - e_{11}^{\bar 1})\mu_{p+k}= \sqrt{p}\mu_{p+k}.
\label{actionmu}
\eea
Just as the basis $\La_k$, $\chi_l$ could be represented
by $(2\times 1)$-arrays of polynomials
in a variable, the same holds for the present basis.
Let us consider
\beq
\mu_k =\left(\begin{array}{c} x^k \\ 0 \end{array}\right),
\qquad
\mu_{p+k}=\left(\begin{array}{c} 0 \\ x^k \end{array}\right),
\qquad k=0,1,\ldots,p-1.
\eeq
When expressed in this basis, the Lie superalgebra will
have a realization preserving the space
\beq
\left(\begin{array}{c} {\cal P}(p-1) \\ {\cal P}(p-1) \end{array}\right).
\label{Pp2}
\eeq
Following from the action given in~(\ref{actionmu}), 
this realization reads~:
\bea
&&b^- = -x^2{d\over dx} + (p-1)x +\si_-, \qquad
b^+ = {d\over dx} + \si_+, \nn\\
&& e_{00}^{\bar 0} - e_{11}^{\bar 0}= 2 x{d\over dx} +1-p-\si_3, \qquad
e_{00}^{\bar 0} + e_{11}^{\bar 0}=p, \nn\\
&& f^- = \sqrt{p} \si_-, \qquad
f^+ = \sqrt{p} \si_+, \qquad
e_{00}^{\bar 1} - e_{11}^{\bar 1}= -\sqrt{p}\si_3, 
\label{real2}\\
&& e_{00}^{\bar 1} + e_{11}^{\bar 1}= {1\over \sqrt{p}}
(-2x{d\over dx}\si_3 +1 +(p-1)\si_3 +2{d\over dx}\si_-+
2(p-1)x\si_+ -2x^2{d\over dx}\si_+). \nn
\eea
and will be referred to as the second differential realization
of $q(2)$.

\section{The sphaleron model}

In this section, we discuss a (physical) system of two coupled 
equations. In particular, this system will have algebraic
solutions in the representation spaces~(\ref{Pp}) and~(\ref{Pp2}).
Such a system arises in the study of the stability of 
sphalerons~\cite{KM} (i.e.\ unstable classical solutions) 
in the Abelian gauge-Higgs model in 1+1 dimensions. 
The relevant equations read~\cite{BK}~:
\bea
&& (\frac{d^2}{dy^2} + \la - \theta^2k^2sn^2)f(y)-2\theta 
k\; cn\; dn\; g(y)=0, \label{36} \\
&& (\frac{d^2}{dy^2} + \la +1+k^2- (\theta^2+2)k^2sn^2)g(y)-
2\theta k\; cn\; dn\; f(y)=0, \label{37}
\eea
and are considered on the Hilbert space of periodic functions 
over $[0,4K(k)]$ ($K(k)$ is the complete elliptic integral of 
the second type). The three elliptic functions~\cite{Arscott} 
$sn=sn(y,k)$, $cn=cn(y,k)$ and $dn=dn(y,k)$ are periodic with 
respective periods $4K(k)$, $4K(k)$ and $2K(k)$. 
The spectral parameter $\la$ is the mode eigenvalue of
the system while $\theta$ stands for the mass ratio $2M_H/M_W$, 
$M_H$ and $M_W$ being respectively the masses of the Higgs 
and gauge bosons.

Introducing the following new function
\beq
W(y) \equiv \frac{df(y)}{dy} - \theta k\; sn \; g(y)
\label{38}
\eeq
as well as of the change of variables
\beq
x=sn^2(y,k), \label{39}
\eeq
the system (\ref{36})-(\ref{37}) becomes
\bea
&&\Bigl(4x(1-x)(1-k^2x)\frac{d^2}{dx^2}+2(1-2(1+k^2)x+3k^2x^2)
\frac{d}{dx}+\la-k^2\theta^2x\Bigr)W(x)=0, 
\label{40} \\
&&\Bigl(4x(1-x)(1-k^2x)\frac{d^2}{dx^2}+2(-1+k^2x^2)\frac{d}{dx}
+\la-k^2\theta^2x\Bigr)f(x)=-2\sqrt{\frac{(1-x)(1-k^2x)}{x}} W(x). \nn\\
&& \label{41}
\eea
It has been proved~\cite{BK} that this system has algebraic
solutions in a $2p$-dimensional space if
\beq
\theta^2=2p(2p+1)\ \hbox{ or }\ \theta^2=2p(2p-1).
\label{42}
\eeq
This result suggests a connection between this sphaleron model 
and the $q(2)$-representations we are dealing with. More precisely, 
if $\theta^2=2p(2p+1)$, we can put either
\beq
W(x)=P_{p-1}(x)+xQ_{p-1}(x), \qquad f(x)=\sqrt{x(1-x)(1-k^2x)}P_{p-1}(x),
\label{43}
\eeq
where $P_m(x)$ and $Q_m(x)$ stand for polynomials of degree $m$ in $x$, 
or else
\beq
W(x)=\sqrt{(1-x)(1-k^2x)}P_{p-1}(x), \qquad 
f(x)=\sqrt{x}(P_{p-1}(x)+xQ_{p-1}(x)).
\label{44}
\eeq
Under one of these two substitutions, the system of 
equations~(\ref{40})-(\ref{41}) has polynomial solutions for
$P_{p-1}(x)$ and $Q_{p-1}(x)$. 
Indeed, in the case~(\ref{43}), the system of equations becomes
\begin{eqnarray}
&&\Bigl(4x(1-x)(1-k^2x)\frac{d^2}{dx^2}+2(1-4(1+k^2)x+7k^2x^2)\frac{d}{dx}
+\la \nn\\
&&\qquad-k^2(4p^2+2p-6)x\Bigr)P_{p-1}(x) =-2Q_{p-1}(x), \label{45}\\
&&\Bigl(4x(1-x)(1-k^2x)\frac{d^2}{dx^2}+2(5-6(1+k^2)x+7k^2x^2)\frac{d}{dx}
+\la-4(1+k^2)\nn\\
&&\qquad -k^2(4p^2+2p-6)x\Bigr)Q_{p-1}(x)
=\Bigl((8k^2x-4(1+k^2))\frac{d}{dx}-6k^2\Bigr)P_{p-1}(x).
\label{46}
\end{eqnarray}
The differential operators of (\ref{45})-(\ref{46}) map any element
$\left(\begin{array}{c}  P_{p-1}(x) \\  Q_{p-1}(x) \end{array}\right)$
of the space~(\ref{Pp2}) into an element of the same space. 
Thus~(\ref{45})-(\ref{46}) reduces to an algebraic eigenvalue system
for $\la$. The differential operator can be written as
\begin{eqnarray}
&&\Delta_{(\ref{43})}+\la= 4x\frac{d^2}{dx^2}-4(1+k^2)x^2\frac{d^2}{dx^2}+
4k^2x^3\frac{d^2}{dx^2}+(6-10(1+k^2)x+14k^2x^2)\frac{d}{dx} \nonumber \\
&&\quad +(-4+2(1+k^2)x)\frac{d}{dx}\si_3+(-4p^2-2p+6)k^2x
-2(1+k^2)+2(1+k^2)\si_3 \nonumber \\
&&\quad +2\si_+-6k^2\si_-+(4(1+k^2)-8k^2x)\frac{d}{dx}\si_- +\la.
\label{47}
\end{eqnarray}
Since this operator leaves the space of polynomials~(\ref{Pp2})
invariant, we might expect that it can be expressed in terms of the 
$q(2)$-generators realized as in the so-called second realization
(i.e.\ as in~(\ref{real2})). We actually have
\begin{eqnarray}
&&\Delta_{(\ref{43})}+\la=2(e^{\bar0}_{00}-e^{\bar0}_{11})b^+ 
-\frac{2}{\sqrt{p}}(e^{\bar0}_{00}-e^{\bar0}_{11})f^+
-2k^2(e^{\bar0}_{00}-e^{\bar0}_{11})b^- \nonumber \\
&&+\frac{2k^2}{\sqrt{p}}(e^{\bar1}_{00}-e^{\bar1}_{11})b^-
-(1+k^2)(e^{\bar0}_{00}-e^{\bar0}_{11})^2
+\frac{2}{\sqrt{p}}b^+(e^{\bar1}_{00}-e^{\bar1}_{11})\nonumber \\
&&-\frac{6}{p}f^+(e^{\bar1}_{00}-e^{\bar1}_{11})
+(1+k^2)\frac{1}{\sqrt{p}}(e^{\bar1}_{00}-e^{\bar1}_{11})
(e^{\bar0}_{00}-e^{\bar0}_{11})-\frac{2k^2}{\sqrt{p}}
(e^{\bar0}_{00}-e^{\bar0}_{11})f^-\nonumber \\
&&+\frac{2k^2}{p}(e^{\bar1}_{00}-e^{\bar1}_{11})f^-
+4(1+k^2)\frac{1}{\sqrt{p}}b^+f^--4(1+k^2)\frac{1}{p}f^+f^-
-2k^2(1-p)\frac{1}{\sqrt{p}}f^- \nonumber \\
&&+2(p+2)b^+-2(p-1)\frac{1}{\sqrt{p}}f^+-6k^2pb^-
-(1+k^2)(2p+1)(e^{\bar0}_{00}-e^{\bar0}_{11}) \nonumber \\
&&+(1+k^2)\sqrt{p}(e^{\bar1}_{00}-e^{\bar1}_{11})-p(p+1)(1+k^2)+\la.
\label{48}
\end{eqnarray}
The same result holds for the case~(\ref{44}) where we obtain
\begin{eqnarray}
&&\Delta_{(44)}+\la=2(e^{\bar0}_{00}-e^{\bar0}_{11})b^+ 
-\frac{2}{\sqrt{p}}(e^{\bar0}_{00}-e^{\bar0}_{11})f^+
-2k^2(e^{\bar0}_{00}-e^{\bar0}_{11})b^- \nonumber \\
&&+\frac{2k^2}{\sqrt{p}}(e^{\bar1}_{00}-e^{\bar1}_{11})b^-
-(1+k^2)(e^{\bar0}_{00}-e^{\bar0}_{11})^2+\frac{2}{\sqrt{p}}b^+
(e^{\bar1}_{00}-e^{\bar1}_{11})\nonumber \\
&&-\frac{6}{p}f^+(e^{\bar1}_{00}-e^{\bar1}_{11})
+(1+k^2)\frac{1}{\sqrt{p}}(e^{\bar1}_{00}-e^{\bar1}_{11})
(e^{\bar0}_{00}-e^{\bar0}_{11})-\frac{2k^2}{\sqrt{p}}
(e^{\bar0}_{00}-e^{\bar0}_{11})f^-\nonumber \\
&&+\frac{2k^2}{p}(e^{\bar1}_{00}-e^{\bar1}_{11})f^-
+4(1+k^2)\frac{1}{\sqrt{p}}b^+f^--4(1+k^2)\frac{1}{p}f^+f^-
+2k^2\sqrt{p}f^- \nonumber \\
&&+2(p+2)b^+-2\sqrt{p}f^+-6k^2pb^--(1+k^2)(2p+1)(e^{\bar0}_{00}
-e^{\bar0}_{11}) \nonumber \\
&&+(1+k^2)\frac{1}{\sqrt{p}}(p+1)(e^{\bar1}_{00}-e^{\bar1}_{11})
-p(p+1)(1+k^2)+\la. \label{49}
\end{eqnarray}
In the case that $\theta^2=2p(2p-1)$, we can consider either
\beq
W(x)=\sqrt{x}Q_{p-1}(x), \qquad f(x)=\sqrt{(1-x)(1-k^2x)}P_{p-1}(x),
\label{50}
\eeq
or else
\beq
W(x)=\sqrt{x(1-x)(1-k^2x)}Q_{p-2}(x), \qquad f(x)=P_p(x).
\label{51}
\eeq
With the substitution~(\ref{50}), the space preserved by the
differential operator is still~(\ref{Pp2}). Acting on an array of 
polynomials $\left(\begin{array}{c}  P_{p-1}(x) \\  Q_{p-1}(x) 
\end{array}\right)$, the equation reduces to an algebraic eigenvalue
equation; using the second realization~(\ref{real2}) one is again able to 
express the differential operator subtended by this physical model 
in terms of the $q(2)$-generators. Explicitely this reads~:
\begin{eqnarray}
&&\Delta_{(\ref{50})}+\la =2(e^{\bar0}_{00}-e^{\bar0}_{11})b^+ -\frac{2}{\sqrt{p}}(e^{\bar0}_{00}-e^{\bar0}_{11})f^+
-2k^2(e^{\bar0}_{00}-e^{\bar0}_{11})b^- \nonumber \\
&&+\frac{2k^2}{\sqrt{p}}(e^{\bar0}_{00}-e^{\bar0}_{11})f^-
+\frac{2k^2}{\sqrt{p}}(e^{\bar1}_{00}-e^{\bar1}_{11})b^-
-\frac{2k^2}{p}(e^{\bar1}_{00}-e^{\bar1}_{11})f^-\nonumber \\
&&-(1+k^2)(e^{\bar0}_{00}-e^{\bar0}_{11})^2+\frac{2}{\sqrt{p}}b^+
(e^{\bar1}_{00}-e^{\bar1}_{11})\nonumber \\
&&-\frac{6}{p}f^+(e^{\bar1}_{00}-e^{\bar1}_{11})
+(1+k^2)\frac{1}{\sqrt{p}}(e^{\bar1}_{00}-e^{\bar1}_{11})
(e^{\bar0}_{00}-e^{\bar0}_{11}) \nonumber \\
&&+2pb^++\frac{2}{\sqrt{p}}(3-p)f^+-2k^2(3p-2)b^-
+2k^2(3p-2)\frac{1}{\sqrt{p}}f^-\nonumber \\
&&+(1+k^2)(-2p+1)(e^{\bar0}_{00}-e^{\bar0}_{11}) 
-(1+k^2)(1-p)\frac{1}{\sqrt{p}}(e^{\bar1}_{00}-e^{\bar1}_{11})\nonumber \\
&&-p(p-1)(1+k^2)+\la.
\label{52}
\end{eqnarray}
The context for the substitution~(\ref{51}) is slightly different, 
so it deserves more attention. This time, the differential operator 
coming from the system~(\ref{40})-(\ref{41}) acts on an element
$\left(\begin{array}{c}  P_{p}(x) \\  Q_{p-2}(x) \end{array}\right)$ 
from the space~(\ref{Pp}). Since also this space is a representation 
space for $q(2)$, as we have proved in the previous section, we can again
expect that the differential operator can be written in terms of the 
$q(2)$-generators. This is indeed the case when using the first 
differential realization of $q(2)$ as given in~(\ref{real1}).
There comes
\begin{eqnarray}
&&\Delta_{(\ref{51})}+\la=2k^2b^+(e^{\bar0}_{00}-e^{\bar0}_{11}) 
-k^2f^+(e^{\bar1}_{00}+e^{\bar1}_{11})-\frac{1}{\sqrt{p}}b^-
(e^{\bar1}_{00}+e^{\bar1}_{11}) \nonumber \\
&&-2(e^{\bar0}_{00}-e^{\bar0}_{11})b^-+\frac{1}{\sqrt{p}}k^2b^+
(e^{\bar1}_{00}+e^{\bar1}_{11}) +4(1+k^2)b^+b^-\nonumber \\
&&+f^-(e^{\bar1}_{00}+e^{\bar1}_{11})+\frac{1}{2}(1+k^2)
(e^{\bar1}_{00}-e^{\bar1}_{11})(e^{\bar1}_{00}+e^{\bar1}_{11})
-\frac{1}{2\sqrt{p}}(1+k^2)(e^{\bar0}_{00}
-e^{\bar0}_{11})(e^{\bar1}_{00}+e^{\bar1}_{11})\nonumber \\
&&+(2p-1)b^-+k^2\sqrt{p}f^++k^2(-6p+1)b^+-\sqrt{p}f^-
+(1+k^2)(2p+\frac{1}{2})(e^{\bar0}_{00}-e^{\bar0}_{11}) \nonumber \\
&&-\sqrt{p}(1+k^2)(e^{\bar1}_{00}+e^{\bar1}_{11})
-\frac{1}{2}\sqrt{p}(1+k^2)(e^{\bar1}_{00}-e^{\bar1}_{11})
+(-2p^2+p)(1+k^2)+\la.  \label{53}
\end{eqnarray}

We have thus written each of the differential 
operators~$\Delta_{(\ref{43})}$,
$\Delta_{(\ref{44})}$, $\Delta_{(\ref{50})}$ and $\Delta_{(\ref{51})}$
associated with the
sphaleron model in terms of the $q(2)$ generators. 
The Lie superalgebra $q(2)$ acts as a ``spectrum generating 
superalgebra'' for this physical model. 
More precisely both the sets of linear differential operators
playing a role in the sphaleron model, those preserving the vector space
of 2-arrays of polynomials of degrees $p-1$ and $p-1$ on the one hand and
those preserving the vector space of 2-arrays of polynomials of degrees
$p$ and $p-2$ on the other hand, 
correspond to realizations of $q(2)$ and make
the determination of $\lambda$ possible. 
Such a determination is
relatively straightforward due to the fact that the (linear) Lie
superalgebra $q(2)$ has a particularly simple structure, 
much simpler than the algebras used in previous 
papers~\cite{BK,BG} devoted to the calculation
of $\lambda$. Indeed in these papers, the algebra $so(4)$ 
(for $\Delta_{(\ref{43})}$, $\Delta_{\ref{44})}$ and 
$\Delta_{(\ref{50})}$) as well as an associative (non-linear) 
graded algebra
denoted by ${\cal A}(2)$ (for $\Delta_{(\ref{51})}$) 
have been used for such a task and
this required heavy techniques in connection with the 
study~\cite{BG} of the irreps of this ${\cal A}(2)$. 
Such a simplification obtained by considering $q(2)$ instead of
${\cal A}(2)$ leads to the hope of a more direct diagonalization
of the operators connected with ${\cal A}(n)$~\cite{BG}
by using $q(n)$.

\section{The Moszkowski model}

We now turn to the Moszkowski model~\cite{Mos}. This is a two-level 
model, each of the levels being $N$-fold degenerate with $N_a$ 
particles of type $a$ and $N_b$ particles of type $b$. The state 
of each particle is specified by the quantum numbers 
$\si = \pm \frac{1}{2}$ (taking the value $\frac{1}{2}$ in the upper 
level and $-\frac{1}{2}$ in the lower level) and $q$ which refers 
to the particular degenerate state within a given level. 
The corresponding Hamiltonian associated to the model reads~\cite{Mos}
\beq
H_M=c\,(J_0(a)-J_0(b))+V\{\hat J_+,\hat J_-\},
\label{54}
\eeq
where $c$ is the energy difference between the two levels and 
$V$ denotes the interaction strength. In~(\ref{54}), the operators 
$J_0(a), J_{\pm}(a)$ are defined according to
\bea
J_0(a)&=&\frac{1}{2} \sum_q(a^+_{q,\frac{1}{2}}a^-_{q,\frac{1}{2}}-
a^+_{q,-\frac{1}{2}}a^-_{q,-\frac{1}{2}}), \label{55}\\
J_+(a)&=&\sum_q a^+_{q,\frac{1}{2}}a^-_{q,-\frac{1}{2}}, \label{56} \\
J_-(a)&=&\sum_q a^+_{q,-\frac{1}{2}}a^-_{q,\frac{1}{2}}, \label{57}
\eea
where $a^+_{q,\pm \frac{1}{2}}$ ($a^-_{q,\pm \frac{1}{2}}$) denotes 
the creation (annihilation) operator of a particle of type $a$ in the 
state $q$ with $\si = \pm \frac{1}{2}$. Similar definitions hold for $J_0(b)$, 
$J_{\pm}(b)$ and we also have
\beq
\hat J_i = J_i (a) + J_i (b), \qquad i=0, \pm .
\label{58}
\eeq
The operators $J_0(i)$, $J_{\pm}(i)$ ($i=a,b$) satisfy the 
$so(4)\equiv sl(2) \oplus sl(2)$ commutation relations
\bea
[J_0(i), J_{\pm}(j)]&=&\pm \delta_{ij} J_{\pm}(i), \label{59} \\
{}[J_+(i), J_-(j)]&=&2\delta_{ij}J_0(i), \qquad (i,j=a,b). \label{60}
\eea
Because of this $sl(2) \oplus sl(2)$ symmetry of the Moszkowski 
Hamiltonian, we can also expect the $q(2)$ Lie superalgebra to play 
a role within this model. 
Associating the operators $J_i$ and $K_i$ ($i=0,\pm$) of~(\ref{so4})
with the current operators $J_i(b)$ and $J_i(a)$ respectively,
we can rewrite $H_M$ as 
\beq
H_M=c\;(K_0-J_0)+V(\{ K_+,K_-\} +\{ J_+,J_-\} +2J_+K_-+2J_-K_+).
\label{61}
\eeq
According to~(\ref{bJK}), this can be rewritten as
\begin{eqnarray}
H_M&=&c\;(\frac{1}{\sqrt{p}}(e^{\bar1}_{00}-e^{\bar1}_{11})-
\frac{1}{2}(e^{\bar0}_{00}-e^{\bar0}_{11}))\nonumber \\
&&+V(\frac{1}{2}p^2-\frac{1}{2}(e^{\bar0}_{00}-
e^{\bar0}_{11})^2+\sqrt{p}(e^{\bar1}_{00}+e^{\bar1}_{11})).
\label{62}
\end{eqnarray}
Although in principle the Hamiltonian $H_M$ can be diagonalized
using the expression~(\ref{61}) in terms of $so(4)$-generators,
it turns out to be much simpler using the expression~(\ref{62})
in terms of $q(2)$-generators together with the second differential 
realization~(\ref{real2}) of $q(2)$. Then the Hamiltonian
becomes
\begin{eqnarray}
H_M&=&c\;(-x\frac{d}{dx}+\frac{1}{2}(p-1)-\frac{1}{2}\si_3)
+V(-2x^2\frac{d^2}{dx^2}+(2p-4)x\frac{d}{dx} \nonumber \\
&&+p+2\frac{d}{dx}\si_-+2(p-1)x\si_+-2x^2\frac{d}{dx}\si_+).
\label{63}
\end{eqnarray}
Considering the action of this on the representation space~(\ref{Pp2}),
or equivalently, the action~(\ref{actionmu}) of~(\ref{62}) on the basis 
vectors~(\ref{mu1})-(\ref{mu2}), leads to an eigenvalue system that
is almost trivial to solve, i.e.~:
\beas
E^+_0&=&pV-(1-\frac{p}{2})c, \\
E^{\pm}_k&=&-2Vk(k-p)+c(\frac{p}{2}-k)\pm \sqrt{V^2p^2+c^2-2(p-2k)Vc},
\quad(k=1,2,\ldots,p-1), \\
E^+_p&=&pV+(1-\frac{p}{2})c.
\eeas
Thus we have recovered the well-known diagonalization of the 
Moszkowski Hamiltonian but by using one of the differential realizations 
of the Lie superalgebra $q(2)$. The latter can then be considered 
as a ``spectrum generating superalgebra" of the Moszkowski model.

\section{The Jaynes-Cummings model}

The well-known Jaynes-Cummings Hamiltonian~\cite{JC} is one of 
the diagonalizable Hamiltonians of quantum optics. It describes 
a two-level atom interacting with a single-mode radiation. 
Under the so-called rotating wave approximation for which only 
real transitions (e.g.\ a photon is absorbed while the electron 
jumps from level~1 to level~2) are taken into account, 
the Jaynes-Cummings Hamiltonian is
\beq
H_{JC}=\omega (a^+a^-+\frac{1}{2})-\frac{1}{2}\omega_0 \si_3 
+g(a^-\si_- + a^+\si_+). \label{JC}
\eeq
Here $\omega$ is the field mode frequency, $\omega_0$ the atomic 
frequency while $g$ is a real coupling constant and, as usual, 
$a^-$ and $a^+$ denote the photon annihilation and creation 
operators, respectively.

In order to determine the spectrum of $H_{JC}$, one can use the 
irreducible representations of the Lie superalgebra $u(1,1)$ as 
shown in~\cite{BRR}. We will prove in this section that the Lie 
superalgebra $q(2)$ can play a similar role and thus be considered 
as a ``spectrum generating superalgebra" for the Jaynes-Cummings 
Hamiltonian. For this purpose, we shall use the basis 
vectors~(\ref{La_k})-(\ref{chi_l}) consisting of the states $\Lambda_k$ 
($k=0,1,\ldots,p$) and $\chi_l$ ($l=1,2,\ldots,p-1$). This time,
however, we shall consider the following realization of these 
basis vectors~:
\beq
\Lambda_k=\left(\begin{array}{c}  px^{p-k} \\  (p-k)x^{p-k-1} 
\end{array}\right),\ (k=0,1,\ldots,p),\qquad 
\chi_l=\left(\begin{array}{c}  0 \\  x^{p-l-1} \end{array}\right),
\ (l=1,2,\ldots,p-1),
\eeq
as opposed to~(\ref{Lachi-x}). This new realization of the basis states 
leads to a third differential realization of the $q(2)$-generators given by
\begin{eqnarray}
&&b^-=-x^2\frac{d}{dx}+(p-1)x+x\si_3 +\si_- , \qquad 
b^+=\frac{d}{dx},\nonumber \\
&&e^{\bar 0}_{00}-e^{\bar0}_{11}=2x\frac{d}{dx}+1-p-\si_3,\qquad
 e^{\bar 0}_{00}+e^{\bar0}_{11}=p, \nonumber \\
&&f^-=\sqrt{p}(x\si_3+\si_--x^2\si_+), \qquad
 f^+=\sqrt{p}\si_+, \nonumber \\
&&e^{\bar 1}_{00}-e^{\bar1}_{11}=\sqrt{p}(-\si_3+2x\si_+), \qquad
 e^{\bar 1}_{00}+e^{\bar1}_{11}=\frac{2}{\sqrt{p}}
 (\frac{p}{2}\si_3+\frac{d}{dx}\si_-). \label{69}
\end{eqnarray}
It has to be noticed that the realization of the $sl(2)$ subalgebra 
generated by $b^-, b^+$ and $e^{\bar 0}_{00}-e^{\bar0}_{11}$ as defined 
in~(\ref{69}) coincides with the one performed in~\cite{Zhdanov}, but 
with other arguments. 
Taking in the Hamiltonian~(\ref{JC}) the realization
\beq
a^+=x \; , \qquad a^-=\frac{d}{dx},  \label{70}
\eeq
we can express $H_{JC}$ as 
\begin{eqnarray}
H_{JC}&=&\frac{\omega}{2}(e^{\bar 0}_{00}-e^{\bar0}_{11})
+\frac{1}{2}p\omega +\frac{g}{2}\sqrt{p}(e^{\bar 1}_{00}
+e^{\bar1}_{11}) \nonumber \\
&&+\frac{1}{2}\frac{g}{\sqrt{p}}(e^{\bar 1}_{00}-e^{\bar1}_{11})
+\frac{1}{2}(\omega_0-\omega+g(p-1))\si_3.
\label{72}
\end{eqnarray}
From this equation it is clear that the $q(2)$ superalgebra is a 
``spectrum generating superalgebra" of the Jaynes-Cummings Hamiltonian 
{\em provided} the detuning $\Delta (\equiv \omega-\omega_0)$ satisfies
\beq
\Delta = g (p-1).
\eeq
Suppose this is the case. Then the action of~(\ref{72}) on the basis
elements $\La_k$ and $\chi_l$ follows from~(\ref{bLachi}) and~(\ref{eLachi}).
In fact, $\La_0$ and $\La_p$ are directly eigenvectors of $H_{JC}$
(with the eigenvalues $E^+_0$ and $E_p^+$ respectively),
whereas the other eigenvectors are simple linear combinations
of $\La_k$ and $\chi_k$ ($k=1,2,\ldots, p-1$). Thus it
is straightforward to recover the Jaynes-Cummings spectrum i.e.
\beas
E_0^+&=&\omega p + \frac{1}{2}(p+1)g, \\
E_k^{\pm}&=&\omega (p-k) \pm g\sqrt{\frac{1}{4}p^2+\frac{1}{2}p
+\frac{1}{4}-k}, \qquad (k=1,2,\ldots,p-1), \\
E_p^+&=&\frac{1}{2}(p-1)g,
\eeas
where the positive integer $p$ is arbitrary.

\section*{Acknowledgements}
The authors would like to thank Y.\ Brihaye for some useful
comments and for pointing out Reference~\cite{BG}.

\end{document}